# Bio-inspired polymers with polaritonic properties from visible to infrared: a material playground to mimic purple bacteria light-harvesting resonators.

Samuel Thomas Holder[1], Carla Estévez-Varela[2], Isabel Pastoriza-Santos[2], Martin Lopez-Garcia[3], Ruth Oulton[1] and Sara Núñez-Sánchez[2]

1 Quantum Engineering Technology Labs, University of Bristol, Bristol, UK
2 CINBIO, Universidade de Vigo, 36310 Vigo, Spain
3 International Iberian Nanotechnology Laboratory, Braga, Portugal

## ABSTRACT

Light-harvesting complexes in natural photosynthetic systems, such as those in purple bacteria, consist of photo-reactive chromophores embedded in densely packed "antenna" systems organized in well-defined nanostructures. In the case of purple bacteria, the chromophore antennas are composed of natural J-aggregates such as bacteriochlorophylls and carotenoids. Inspired by the molecular composition of such biological systems, we create a library of organic materials composed of densely packed J-aggregates in a polymeric matrix, in which the matrix mimics a protein scaffold. This library of organic materials shows polaritonic properties which can be tuned from the visible to the infrared by choice of the model molecule. Inspired by the molecular architecture of the light-harvesting complexes of *Rhodospirillum molischianum* bacteria, we study the light-matter interactions of J-aggregate-based nanorings with similar dimensions to the analogous natural nanoscale architectures. Electromagnetic simulations show that these nanorings of J-aggregates can act as resonators, with subwavelength confinement of light while concentrating the electric field in specific regions. These results open the door to bio-inspired building blocks for all-organic metamaterials while offering a new perspective on light-matter interactions at the nanoscale in densely packed organic matter in biological organisms including photosynthetic organelles.

## Introduction

Natural biological systems have evolved complex structures at the molecular level which have been carefully optimised for specific functions. This complexity can be observed, for example, in membrane proteins where chemical selectivity is achieved through specific molecular key/site pairs[1] or, in the arrangement of chromophore molecules in photosynthetic complexes (PCs)[2]. These PCs are composed of supramolecular chromophores contained in well-defined nanostructures where proteins play the role of responsive scaffolds.[3–7] These molecular arrangements have evolved to optimise light capture and exciton transport between chromophores, with proteins modulating energy pathways between them depending on sun irradiation conditions.[8] A significant amount of prior work has focussed on the investigation of PCs as absorbers and emitters by ultrafast spectroscopy and quantum modelling.[9,10] However, these studies neglect the optical properties of the photosynthetic matter and hence, the potential photonic modes supported by the molecular nanostructures. In this work, inspired by the composition of the natural photosynthetic nanostructures of light-harvesting complexes (LHCs) of purple bacteria, we build up a library of bio-inspired organic matter which we use to model the light-matter interaction of photosynthetic nanostructures.

LHCs of purple bacteria are composed of organic matter with a high concentration of dye molecules without presenting quenching (0.5-0.6 M)[11,12] They show annular structures composed of densely packed J-aggregates of π- conjugated organic molecules (carotenoids, bacteriochlorophylls, etc)[13–15] These supramolecular assemblies form a natural J-aggregate with an exciton delocalised across the monomers. Previous research on light-trapping and exciton transport strategies in natural systems typically analyses photosynthetic matter as an effective dielectric medium with a dispersive value of the refractive index. This definition of a macroscopic effective index of the membranes has furthermore been used in advanced models that describe proteins within photosynthetic complexes.[4,16] While this may be an appropriate approximation for materials mainly composed of collagen or cellulose, photosynthetic LHCs nanostructures of purple bacteria are composed of densely packed J-aggregates with strong absorptions and delocalized excitons which can promote a significant modification of the local refractive index, with important consequences for how light interacts with these organic nanostructures.[11,12,17,18] Our previous work demonstrates that thin films composed of densely packed J-aggregates within a polymer have strongly modified optical properties, achieving negative values of the real part of the permittivity and being able to support surface exciton polaritons (SEPs).[19,20] Here we demonstrate that these polaritonic properties are not unique to a single specific molecular compound. On the contrary, the polaritonic character can be achieved with a whole catalogue of molecules across the VIS and NIR spectral range, mirroring the broad range of chromophores appearing in photosynthetic organisms. These polaritonic properties arise from the presence of strong electric dipoles from delocalised Frenkel excitons in the constituent J-aggregates which are embedded within the polymer matrix. As LHCs are composed of densely packed J-aggregates within protein scaffolds we propose that the unusual optical properties required for SEPs may exist in the matter forming LHC nanostructures. Therefore exciton-polariton modes could play a role in the well-defined nanostructures of LHCs[21–23] of purple bacteria, with structural similarities between these natural nanoscale systems and nanoscale metamaterial building blocks.[24,25]

To investigate the physics of polaritons in LHCs of purple bacteria, we study the light-matter interactions of LHC architectures at the nanoscale by electromagnetic simulations, using as models the architecture of the LHC-2 of *Rhodospirillum molischianum* bacteria, together with the optical properties of a library of organic materials composed by artificial J-aggregates.[12,26,27] These simulations revealed that LHC-2 composed of densely packed J-aggregates act as nano-resonators confining the light at the nanoscale. The optical response obtained depends strongly on the polarization of the incident light, revealing that when the incident polarization is contained in the photosynthetic membrane, the electric field is concentrated in the centre and edges of the nanorings, enhancing interaction with the reaction centre and promoting the coupling between closed LHCs. This offers a new perspective on the understanding of light-matter interactions in natural photosynthetic complexes of purple bacteria, demonstrating how supramolecular nanostructures of densely packed J-aggregates can shape and optimise the interaction of light and transport of energy within photosynthetic organelles. Understanding this mechanism could boost light-harvesting efficiency in metamaterial devices through a new family of molecular materials that mimic the carefully optimised molecular arrangement, concentration, and architectural design of natural photosynthetic systems, with J-aggregate and polymer materials forming the building blocks of such a platform.

**Bio-inspired polaritonic library**

In analogy with proteins in the biological systems, we mixed J-aggregates with a water-based polymer which physically separates the J-aggregates giving robustness to the final bulk photosynthetic-mimetic material. We controlled the conformal arrangement of the molecular aggregates to promote self-assembly into J-aggregates by increasing the dye concentration in final dye-polymer water solutions.[28] As a library of J-aggregates, we used commercial water-soluble closed-chain cyanine dyes with optical responses from the visible to the infrared. The structure of the monomers of the four J-aggregates selected for this work is shown in Figure 1a-d: J562, J587, J619 and J798 (see methods for complete name). The Figure 1.a-d shows the extinction of the monomer obtained from the dye solution in ethanol for the four cyanine dyes with peaks at 504 nm (J562), 519 nm (J587), 554 nm (J619) and 664 nm (J562). For a water-based polymer we selected poly (vinyl alcohol) (PVA) with a molecular weight of 85,000-124,000. The four cyanine water solutions (25 mM) were mixed with 6 wt% aqueous PVA solution (3:1 volume ratio) as previously described.[19] As shown in Figure 1.a-d, all four cyanine-PVA solutions show a narrow absorption peak red-shifted with respect to the peak of the monomeric dye that confirms the formation of J-aggregates (measurement protocol in section 1, supplementary). The J-aggregate:PVA films were prepared by spin-coating (10000 rpm) of dye-PVA solutions on glass substrates obtaining a final mass ratio of J-aggregate to PVA in the bulk material of around 1:1.

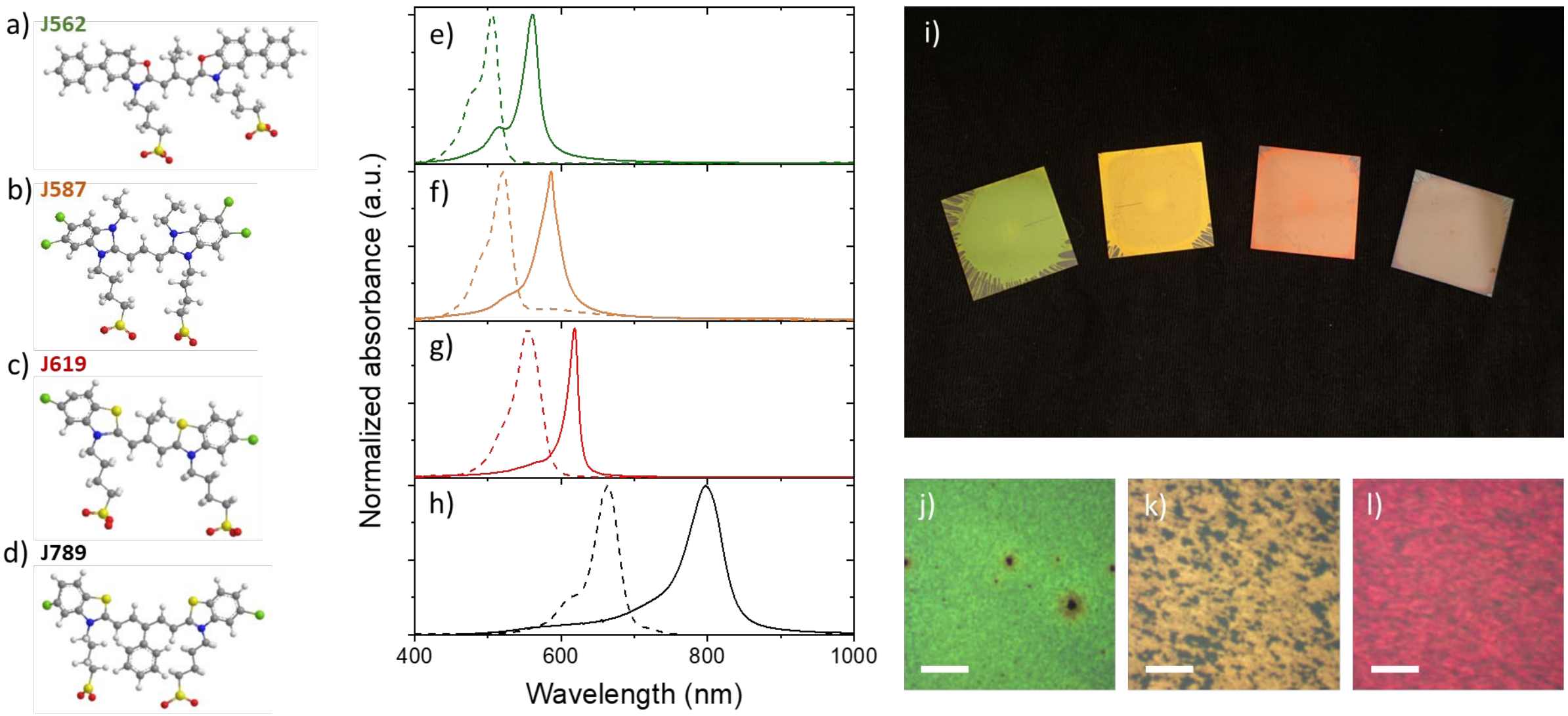

**Figure 1: a-d)** Molecular structure of the monomers of the selected cyanine library for this work. They are named here according to the absorption peak of the J-aggregate conformation (complete name of molecules in methods section). **e-h)** Normalized absorbance of monomeric dye solutions in ethanol (doted lines, 100 μM) and J-aggregate:PVA mixtures (solid line) for the J-aggregates **e)** J652, **f)** J587, **g)** J619 and **h)** J789. **i)** Picture of the J-aggregate:PVA films on top of black cardboard. From left to the right: J562, J587, J619 and J789 films. Optical microscope images of a 250 μm square area for dye **(j)** J562 **(k)** J587 and **(l)** J619 measured in reflectance microscopes under Köhler Illumination configuration. The scale bar is 50 μm.

Figure 1.i shows a picture of the four samples obtained. All the samples show a metallic lustre, each in a different spectral range, observed by the naked eye and through a microscope (Figure 1.j-l). This high reflectance has a well-defined, vivid colour, because it is associated with a narrow region of negative real electric permittivity created by the intense absorption of the densely packed J-aggregates making up each film[19]. While metals generally have negative real electric permittivity in a broad spectral range, and therefore are reflective in a broad spectral range, these J-aggregate based materials exhibit this behaviour in a narrow spectral range, located at shorter wavelengths than the J-aggregate absorption, giving them a vividly coloured reflectance. Reflectance close to 60% is achieved for all four samples (see Figure

2.a). Besides, it is independent of the angle of incidence, demonstrating that it is inherent to the optical properties of the material and is not due to the creation of a Fabry-Perot cavity. The J-aggregate:PVA films were analysed by Atomic Force Microscopy (supplementary section 2). Table 1 shows the thickness and roughness of the different J-aggregate:PVA films. In general, the films are homogeneous on a 100nm length scale. Regardless the dyes, the J-aggregate:PVA films present a roughness comparable to thatof a metal film obtained by thermal evaporation (1-2 nm)[29] being the films with the largest roughness those obtained with J562 and probably due to randomly orientated bundles of J-aggregates at the film surface. For example, the J562 film with the largest roughness shows grain-like features (Figure S2.a). Further work is required to investigate potential domain structure within J-aggregate:PVA materials: in this work we have assumed that the films are optically homogenous.

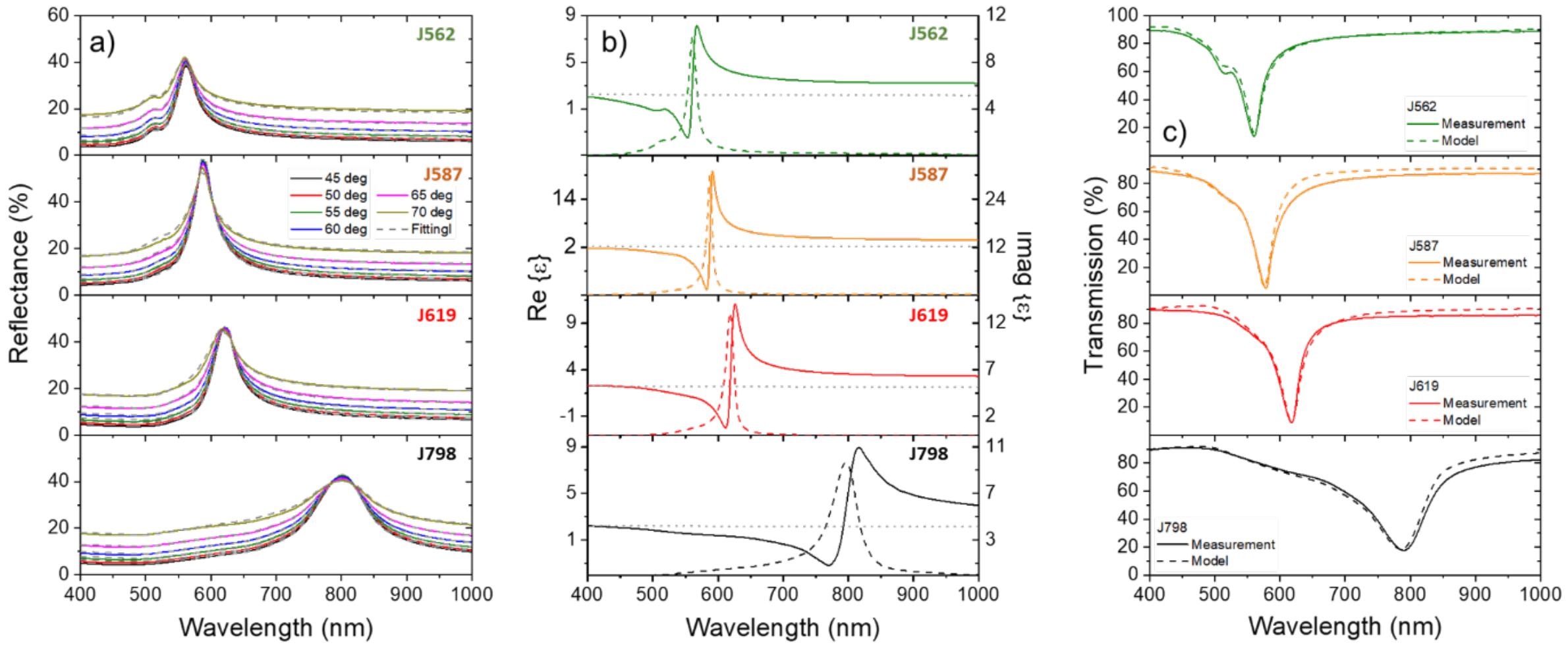

**Figure 2: (a)** Measured (solid lines) and fitted data (dashed lines) of un-polarised reflectance of J-aggregate:PVA films at an angle of incidence of 45, 50, 55, 60, 65 and 70 degrees. **b)** Real (solid line) and imaginary (dashed line) parts of the permittivity for the different J-aggregate:PVA films as indicated obtained by fitting the unpolarised reflectance data only. Real part of the permittivity of PVA from published data (grey dotted line)[30]. **c)** Experimental and modelled transmission through the polymer films using the optical properties obtained from reflectance and the thickness from AFM. Panels from top to bottom: J562, J587, J619 and J789 films.

The optical properties of the films were estimated by fitting the unpolarised reflectance as a function of the incidence angle. The samples were modelled as a stack of two smooth layers: a thin J-aggregate:PVA film on top of a semi-infinite glass substrate. The thickness of the J-aggregate:PVA film was fixed to the measured value for each sample (Table 1).The resulting real and imaginary parts of the relative electric permittivity of each film are shown in Figure 2.b. To confirm the measured optical properties of each film, we measured the transmission spectrum through each sample and compared the experimental values with the modelled transmission spectrum (see Figure 3.c). The good agreement between measured and modelled data demonstrates the accuracy of the obtained optical properties of each material.

We would like to remark that the position of the minimum of the transmissions matches the absorption peaks of the J-aggregate:PVA mixtures (Figure 1e-h) confirming the presence of only J-aggregates in the spin-coated J-aggregate:PVA films. Each J-aggregate:PVA material achieves negative values of the real part of the relative electric permittivity. We have defined the polaritonic band of the material as the wavelength range where the real part of the permittivity is lower than -1 (see table 1), where we expect the materials to support SEPs in the film-air interface. We would like to remark on the low values of real electric permittivity achieved, for example -8.3 in the case of J587 (see Table 1).

**Table 1**.- Roughness ($\sigma$), thickness, polaritonic band, full half maximum width (FHMW) and minimum value of real part of permittivity ($\varepsilon_1$(min)) of the J-aggregate:PVA films obtained with the different dyes.

| **Dye** | $\sigma$ (nm) | **Thickness** (nm) | **Polaritonic band*** | **FHMW** (meV)* | $\varepsilon_1$ (min) |
|---|---|---|---|---|---|
| **J562** | 5.1 ± 0.3 | 31 ± 1.3 | 548 – 557 nm | 55 | -1.5 |
| **J587** | 3.2 ± 0.3 | 23 ± 0.8 | 552 – 587 nm | 36 | -8.3 |
| **J619** | 2.1 ± 0.1 | 38 ± 0.4 | 600 – 615 nm | 49 | -2.2 |
| **J798** | 2.0 ± 0.1 | 33 ± 0.7 | 763 – 775 nm | 92 | -1.2 |

*The polaritonic band is defined as the wavelength regions where the J-aggregate:PVA films are polaritonic (real part of permittivity < -1 ). Full Half Maximum Width (FHMW) is the FHMW of the imaginary part of the permittivity.

Based on the measured optical properties, each J-aggregate:PVA film is expected to support SEPs at the air-polymer interface within the polaritonic band. The top panels of figure 3 show the p-polarised reflectance from 400 nm to 850 nm as a function of the angle for the four J-aggregate:PVA films measured under Kretschmann prism-coupling configuration by Fourier imaging spectroscopy (detailed description in S3). A dip in the p-polarised reflectance is observed for the four samples at larger angles than the critical angle for glass (41 degrees). This dip in reflectance is observed in a narrow wavelength range corresponding to the polaritonic band specific for each constituent J-aggregate. The observed dips are broad in angle, suggesting that the SEPs supported by our J-aggregate:PVA films are affected by losses associated to the imaginary part of the permittivity. However, we would like to remark on the contrast between p- and s-polarisation measured for all samples, showing efficient mode coupling at central wavelengths of the polaritonic band (bottom panels of Figure 3).

In summary, all four J-aggregate:PVA materials show a wavelength-specific polaritonic response depending on the optical response of the constituent J-aggregate. Therefore, the negative permittivity region can be selected through the characteristics of the constituent J-aggregate, creating a library of plasmonic-like materials, with the potential to expand to a much broader range of molecules including cyanines and porphyrins. Table 1 shows the wavelength ranges where the J-aggregate:PVA materials exhibit real relative electric permittivity less than -1, where SEPs were excited. Table 1 also shows the full-width half maximum (FWHM) of the imaginary part of the permittivity and the minimum value reached. Note that broader optical resonances can yield negative permittivity if the dye in question reaches a sufficiently high density.

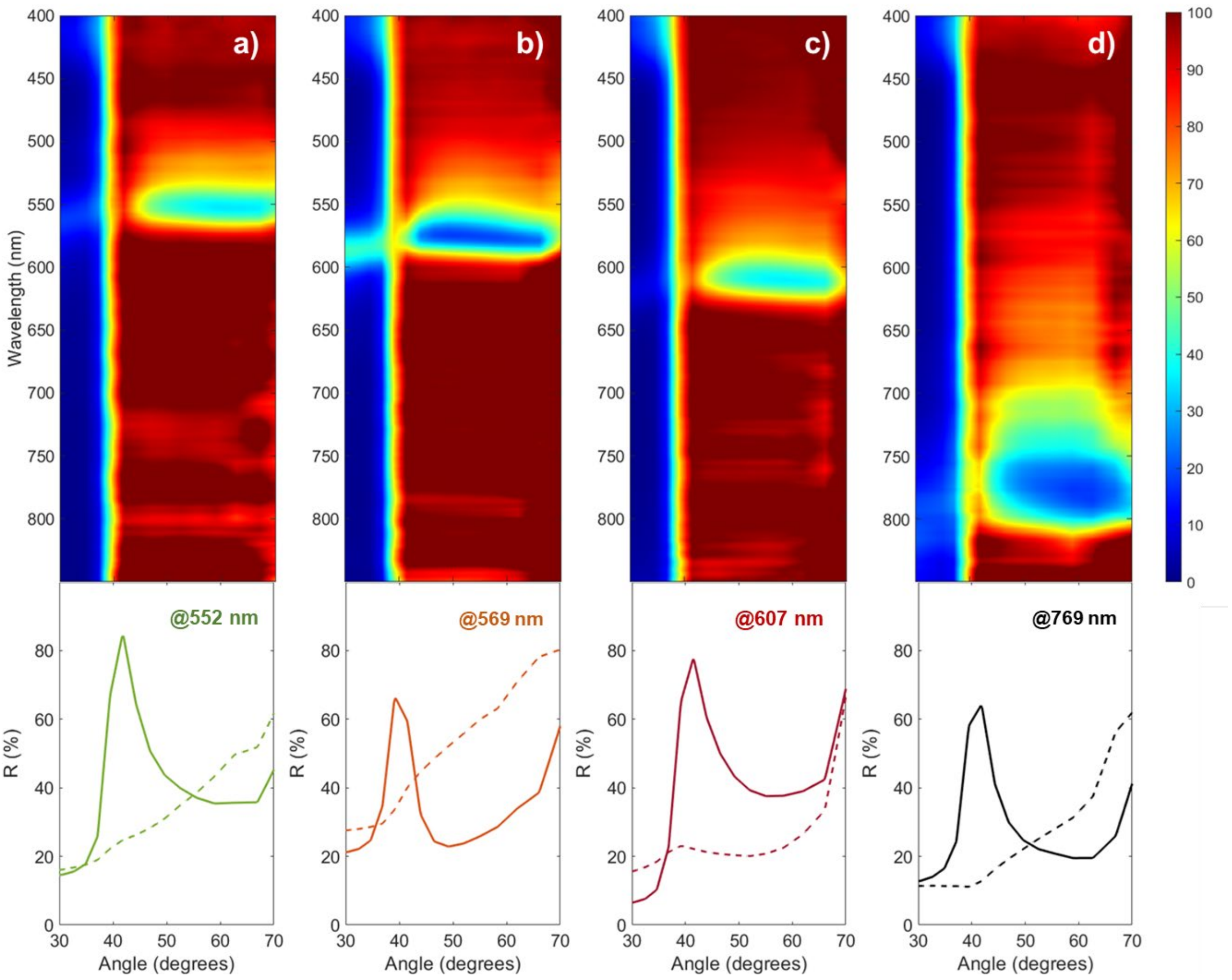

**Figure 3:** Measured p-polarised reflectance from J-aggregate:PVA films containing **(a)** J562, **(b)** J587, **(c)** J619 and **(d)** J798 J-aggregates. The bottom panels show the p-polarised and s-polarised reflectance as a function of the incident angle at the central wavelength of the polaritonic band for each film: **a)** 552 nm for J562, **b)** 569 nm for J587, **c)** 607 nm for J619 and **d)** 769 nm for J789.

### Electromagnetic modelling of purple bacteria light-harvesting complexes

Considering that natural photosynthetic nanostructures of LHCs of purple bacteria contain densely packed J-aggregates, we hypothesise that these biological structures may also support SEPs, specifically local surface excitonic resonances (LSERs). These modes would be analogous to local surface plasmon resonances in plasmonic metamaterials, and we model the nanophotonic response of individual nanorings of densely packed dyes mimicking natural LHCs architectures as potential building blocks of an all-organic metamaterial platform, using the obtained optical properties of J-aggregate:PVA materials. Note that in comparison with natural LHCs, our model does not seek to exactly replicate effects such as optical losses, molecular disorder or coupling strength between constituents monomers.[31] Instead, we are drawing an analogy between metamaterials and natural architectures of densely packed J-aggregates with important implications for the nanophotonics of the respective systems, including how light may be managed in natural photosynthetic supramolecular nanostructures through polaritonic modes mediating exciton transport or energy transfer.[32,33]

There are a large variety of dye architectures described across the different photosynthetic systems in lamellar membranes, spherical vesicles or tubular structures.[34] But the most common feature of photosynthetic species of purple bacteria is a photosynthetic unit containing the reaction centre surrounded by peripheral dyes and proteins creating closed and open loop nanostructures. The size and dye distributions between species can change, but in the case of green and purple bacteria, LHCs show ring structures with a total external diameter varying between 5 nm[5] to more than 10 nm with some variations depending on the species.[5,15,35,36] In this work we have selected as an example the LHC-2 ring architecture of lamellar membranes of *Rhodospirillum molischianum* bacteria which are distributed randomly on top of a lamellar flat membrane.[35] Therefore, our artificial model of the LHC-2 architecture is a ring with an outer and inner diameter of 9 nm and 3.1 nm respectively, and a height of 5 nm (see Figure 4.a), composed of a J-aggregate:PVA material as analysed above. This model will illustrate the nanophotonics of bio-mimetic ring structures containing densely packed J-aggregates, whether the ring structure consists of natural J-aggregates in a protein scaffold or synthetic cyanine J-aggregates in a PVA matrix.

The optical response and distribution of local electric fields have been estimated by Finite Difference Time Domain simulations using commercial software (Lumerical-Ansys). Rings of five different materials were simulated: the four J-aggregate:PVA materials studied above, and a fifth ring of PVA-only with no dyes as a reference. The methodology and parameters used for the simulations are described in detail in section 5 of the Supplementary. Figure 4.b shows the extinction cross-sections obtained for all J-aggregate:PVA nanorings and the reference. The reference shows a flat null response from 400 to 900 nm. However, the J-aggregate:PVA nanorings showed a clear extinction peak at shorter wavelengths than the absorption of the constituent J-aggregate but within the polaritonic band of each material (Table 1 and Figure S8), which is associated with LSERs. Importantly the extinction peak of the nanoring is centered at shorter wavelengths (higher energies) than the absorption peak of the monomeric dye: this implies that rings may be capable of harvesting light at higher energies than the absorption peak of constituent dye. This means that light-harvesting could be enhanced by polaritonic resonances associated with natural metamaterial nanostructures in the spectral range in which the constituent dyes themselves are less efficient at absorbing light directly. Furthermore, the proximity of the polaritonic resonance to the maximum of the dye absorption will increase the optical density of states at these wavelengths enhancing the absorption and emission rate of the constituent dyes in the nanorings.

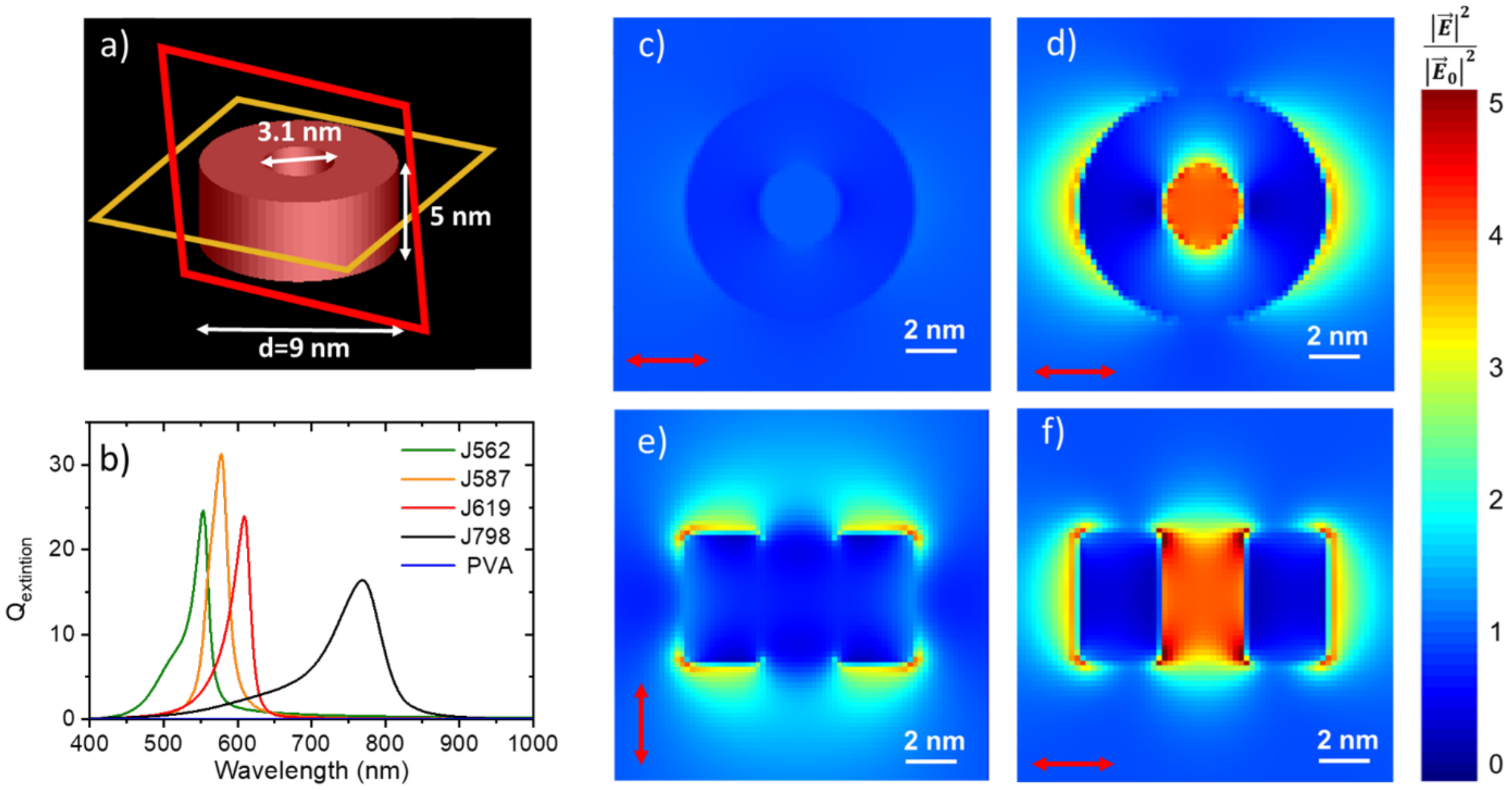

**Figure 4.- a)** Bioinspired architecture of LHC of *Rhodospirillum molischianum* bacteria including the dimensions of the nanostructure. The red rectangle (XZ plane) and yellow rectangle (XY plane) indicate planes crossing the centre of the nanoring, where the distribution of the electric field is estimated at the peak wavelength of the extinction. **b)** Normalized extinction efficiency cross section of simulated LH2 nanorings composed of only pure PVA and J-aggregate:PVA materials obtained via combination of J562, J587, J619 or J798 with PVA. The Local Surface Excitonic Resonances are centred at 553 nm, 578 nm, 609 nm and 768 nm for J562, J587, J619, and J798 based materials, respectively. **(c-d)** Electric field intensity distribution at 609 nm in the planes XY **(e-f)** XZ crossing the centre of the nanoring. Red arrows indicate polarization, with light propagating perpendicular to the plane shown in each figure. **c)** PVA-only nanoring, **(d-f)** J619 J-aggregate:PVA nanoring.

We now investigate the electric field distribution associated with the LSER excitations of the J-aggregate:PVA LHC analogue nanorings, to provide insight into how the structure may control electromagnetic fields to improve light-harvesting. The distribution of the electric field is key to demonstrating firstly light confinement and secondly, how the electric field is distributed, enhancing light-matter interactions in specific areas and coupling with other molecular entities. The LHCs of *Rhodospirillum molischianum* bacteria are distributed in the photosynthetic organelle within a lamellar photosynthetic membrane. Therefore, we considered two illumination incident conditions, parallel and perpendicular to the lamellar membrane. Figure .4a shows a three-dimensional representation of the modelled LHC nanoring. The yellow plane contains the lamellar membrane while the red plane is perpendicular to it. We would like to remark that in natural photosynthetic membranes of *Rhodospirillum molischianum* bacteria, LHCs are closely packed within the membrane with a reaction centre located at the centre of some LHC nanorings.

Here we focus on LSERs of the J619 nanoring (similar results are obtained for J562 J587 and J798 J-aggregate:PVA materials, see Figures S10-S12). The Figure 4 (c-f) shows the electric field intensity distribution at 609nm (LSER peak for J619) for the reference nanoring (PVA only, Figure 4.c) and a nanoring composed of J619 J-Aggregate:PVA material (Figure 4.d-f). In the case of the reference nanoring, there is no local field enhancement, with a field intensity distribution close to one across the nanostructure (see Figure 4.c). However, the electric field distribution is drastically different for the nanoring made of J619 J-aggregate:PVA. In this case, when the incident polarization is in the plane of the lamellar membrane (Figure 4.d and 4.f), the electric field is concentrated at the centre and edges of the nanoring in the direction of the polarization. Note that the reaction centre of a natural LHC is usually located at the centre of the nanoring structure. Therefore, our results suggest that the presence of J-aggregates in the natural nano-architecture may modify how light and excitons interact with the molecules of the reaction centre. Furthermore, the high electric field obtained at the edges of the ring could enhance optical coupling between adjacent nanorings through evanescent fields. As natural membranes are formed by densely packed LHCs, this suggests that the presence of J-aggregates within molecular nanoring architectures could promote exciton transport between nanorings by polaritonic coupling between adjacent LHCs nanorings. Further studies analysing complex distributions of LHCs nanorings are required for a more detailed analysis, though this is out of the scope of this article. Finally, it is remarkable how the optical response is strongly dependent on polarization. While incident polarization in the plane of the lamellar membrane concentrates the electric field in the centre of the nanoring and its lateral edges, when the polarization is perpendicular to the lamellar membrane, the electric field at the centre is zero, and high field intensities are found at the top and bottom interfaces of the nanorings.

## Conclusions

In this article, we have developed a library of densely packed supramolecular bulk materials based on J-aggregates and PVA, inspired by the molecular composition of natural supramolecular nanostructures: LHCs of purple bacteria. Our results show that all of the studied J-aggregate:PVA materials have a strongly dispersive optical response creating highly reflective materials that support surface exciton polaritons from the visible to the infrared. The measured optical properties of these materials were used to explore how light interacts with ringed nanostructures with similar architecture to the natural ringed nanostructures found in the photosynthetic membrane of the purple bacteria *Rhodospirillum molischianum,* but composed of synthetic J-aggregate:PVA materials. All modelled nanostructures showed subwavelength confinement concentrating the optical field in the centre and edges of the nanorings through local surface exciton resonances. This demonstrates how densely packed dyes in synthetic nanostructures, similar to those found in natural systems, can create light confinement and field enhancement through electromagnetic excitonic resonances, similar to plasmonics. The similarities between the optical and excitonic properties of the natural and synthetic materials suggest that exciton polariton resonances could play a role in photosynthesis in purple bacteria. Furthermore, the specific features of these resonances show that they could be exploited to improve light-harvesting efficiency by broadening the wavelength response of constituent dyes while aiding energy transfer to the reaction centre and between nanorings through control and enhancement of the local electric field. Further studies on natural, larger scale distributions of LHCs in photosynthetic organisms are required to determine if evolution has developed nanophotonic strategies to optimize energy capture and exciton transport within photosynthetic organelles using excitonic metamaterials. With this work, we would like to offer a new perspective for the development of novel organic metamaterial platforms and structures, taking inspiration from nature across length scales: from the molecular scale, passing through the mesoscale, towards well-defined large area metamaterial structures.

## SUPPORTING INFORMATION

Description of experimental methodology to measure transmission in PVA:J-aggregate solutions. AFM graphical data. A detailed description of measurement conditions and calibration protocol for polarised reflectance under Kretschmann prism-coupling configuration by Fourier imaging spectroscopy. FDTD simulations: scattering and absorption cross sections and Local Field Intensity distributions in nanorings composed by J-aggregate:PVA materials of the combination of J562, J587, and J798 with PVA

## ACKNOWLEDGEMENTS

S.T.H acknowledges financial support from UK EPSRC grant QuPIC (EP/N015126/1) and UK EPSRC Quantum Engineering Centre for Doctoral Training, University of Bristol. I.P.-S., S. N.-S. and C.E.-V. acknowledge financial support from Xunta de Galicia/FEDER (grants GRC ED431C 2020/09 and GR 2007/085). M.L.G acknowledges the support of the grant NASCADIA, code PTDC/BTA-BTA/2061/2021, funded by FCT of Portugal.

## METHODS

### Chemicals

All dyes were purchased in from Few Chemicals:

- J562: (5-Phenyl-2-[2-[[5-phenyl-3-(4-sulfobutyl)-3H-benzoxazol-2-ylidene]-methyl]-but-1-enyl]-3-(4-sulfobutyl)-benzoxazolium hydroxide.
- J587: (5,6-Dichloro-2-[[5,6-dichloro-1-ethyl-3-(4-sulfobutyl)-benzimidazol-2-ylidene]-propenyl]-1-ethyl-3-(4-sulfobutyl)-benzimidazolium hydroxide.
- J619: (5-Chloro-2-[2-[5-chloro-3-(4-sulfobutyl)-3H-benzothiazol-2-ylidenemethyl]-but-1-enyl]-3-(4-sulfobutyl)-benzothiazol-3-ium hydroxide.
- J798 (: 5-Chloro-2-[5-[5-chloro-3-(4-sulfobutyl)-3H-benzothiazol-2-ylidene]-3-phenyl-penta-1,3-dienyl]-3-(4-sulfobutyl)-benzothiazol-3-ium hydroxide.
- PVA (Poly(vinyl alcohol), Mw 85,000-124,000, 99+% hydrolysed) was purchased in Sigma Aldrich.
- All water solutions were prepared with Milli Q-water.
- All ethanol solutions were prepared with ethanol synthesis grade.

**Preparation of PVA and cyanine water solutions.**

The 6 wt% aqueous PVA solution was prepared under stirring at 90°C for one hour. The 25 mM solution of each dye was made under gentle stirring for several hours until a homogeneous solution of J-aggregates was reached.

**Deposition of J-aggregate:PVA films.**

The J-aggregate:PVA film was obtained by spin-coating at 1000 rpm on top of a cover glass. Previously the deposition, the cover glass substrates were bathed overnight in 37 wt % HCl before rinsing in deionised water and drying with an air hose to obtain a hydrophilic surface.

**Unpolarised reflectance measurements.**

Unpolarised reflectance spectra at six equally spaced angles of incidence between 45 deg and 70 deg were measured using a J.A. Woollam RC2 ellipsometer. Scotch tape was applied to the underside of each sample during measurement to provide an index-matched, rough bottom surface eliminating reflections from the bottom of the sample.

**P-polarised and s-polarised reflectance under Kretschmann prism-coupling configuration**

In this work, we used a tungsten-halogen white light lamp covering UV–vis–NIR spectral range for illumination. The optics consisted of a high numerical aperture Nikon plan APO 100x with NA = 1.45 mounted on a Nikon Ti2 microscope body. A fiber-coupled 2000+ Ocean Optics (Dunedin, USA) spectrometer was used. The prism coupling condition was obtained by illuminating from the glass side of each sample using an immersion oil objective. The calibration of the angular response is described in detail in section S3.

## REFERENCES

(1) Cellular Gatekeepers. *Nat. Struct. Mol. Biol.* **2016**, *23* (6), 463. https://doi.org/10.1038/nsmb.3246.

(2) Blankenship, R. E. *Molecular Mechanism of Photosynthesis*, 2nd Editio.; Jonh Wiley & Sons, 2014.

(3) Jacobs, M.; Lopez-Garcia, M.; Phrathep, O. P.; Lawson, T.; Oulton, R.; Whitney, H. M. Photonic Multilayer Structure of Begonia Chloroplasts Enhances Photosynthetic Efficiency. *Nat. Plants* **2016**, *2* (October), 1–6. https://doi.org/10.1038/nplants.2016.162.

(4) Capretti, A.; Ringsmuth, A. K.; van Velzen, J. F.; Rosnik, A.; Croce, R.; Gregorkiewicz, T. Nanophotonics of Higher-Plant Photosynthetic Membranes. *Light Sci. Appl.* **2019**, *8* (1), 5. https://doi.org/10.1038/s41377-018-0116-8.

(5) Scheuring, S.; Reiss-Husson, F.; Engel, A.; Rigaud, J. L.; Ranck, J. L. High-Resolution AFM Topographs of Rubrivivax Gelatinosus Light-Harvesting Complex LH2. *EMBO J.* **2001**, *20* (12), 3029–3035. https://doi.org/10.1093/emboj/20.12.3029.

(6) Gonçalves, R. P.; Busselez, J.; Lévy, D.; Seguin, J.; Scheuring, S. Membrane Insertion of Rhodopseudomonas Acidophila Light Harvesting Complex 2 Investigated by High Resolution AFM. *J. Struct. Biol.* **2005**, *149* (1), 79–86. https://doi.org/10.1016/j.jsb.2004.09.001.

(7) Scheuring, S.; Gonçalves, R. P.; Prima, V.; Sturgis, J. N. The Photosynthetic Apparatus of Rhodopseudomonas Palustris: Structures and Organization. *J. Mol. Biol.* **2006**, *358* (1), 83–96. https://doi.org/10.1016/j.jmb.2006.01.085.

(8) Romero, E.; Novoderezhkin, V. I.; van Grondelle, R. Quantum Design of Photosynthesis for Bio-Inspired Solar-Energy Conversion. *Nature* **2017**, *543* (7645), 355–365. https://doi.org/10.1038/nature22012.

(9) Fassioli, F.; Dinshaw, R.; Arpin, P. C.; Scholes, G. D. Photosynthetic Light Harvesting: Excitons and Coherence. *J. R. Soc. Interface* **2014**, *11* (92), 20130901. https://doi.org/10.1098/rsif.2013.0901.

(10) Baghbanzadeh, S.; Kassal, I. Geometry, Supertransfer, and Optimality in the Light Harvesting of Purple Bacteria. *J. Phys. Chem. Lett.* **2016**, 1–7. https://doi.org/10.1021/acs.jpclett.6b01779.

(11) Mirkovic, T.; Ostroumov, E. E.; Anna, J. M.; Grondelle, R. Van; Scholes, G. D. Light Absorption and Energy Transfer in the Antenna Complexes of Photosynthetic Organisms. *Chem. Rev.* **2016**, 1–103. https://doi.org/10.1021/acs.chemrev.6b00002.

(12) Brixner, T.; Hildner, R.; Köhler, J.; Lambert, C.; Würthner, F. Exciton Transport in Molecular Aggregates - From Natural Antennas to Synthetic Chromophore Systems. *Adv. Energy Mater.* **2017**, *1700236* (16), 1–33.

https://doi.org/10.1002/aenm.201700236.

(13) Linnanto, J.; Oksanen, J. A. I.; Korppi-Tommola, J. E. I. Exciton Interactions in Self-Organised Bacteriochlorophyll a - Aggregates. *Phys. Chem. Chem. Phys.* **2002**, *4* (13), 3061–3070. https://doi.org/10.1039/b106692g.

(14) Sengupta, S.; Würthner, F. Chlorophyll J-Aggregates: From Bioinspired Dye Stacks to Nanotubes, Liquid Crystals, and Biosupramolecular Electronics. *Acc. Chem. Res.* **2013**, *46* (11), 2498–2512. https://doi.org/10.1021/ar400017u.

(15) Van Rossum, B. J.; Steensgaard, D. B.; Mulder, F. M.; Boender, G. J.; Schaffner, K.; Holzwarth, A. R.; De Groot, H. J. M. A Refined Model of the Chlorosomal Antennae of the Green Bacterium Chlorobium Tepidum from Proton Chemical Shift Constraints Obtained with High-Field 2-D and 3-D MAS NMR Dipolar Correlation Spectroscopy. *Biochemistry* **2001**, *40* (6), 1587–1595. https://doi.org/10.1021/bi0017529.

(16) Castillo, M. A.; Wardley, W. P.; Lopez-Garcia, M. Light-Dependent Morphological Changes Can Tune Light Absorption in Iridescent Plant Chloroplasts: A Numerical Study Using Biologically Realistic Data. *ACS Photonics* **2021**, *8* (4), 1058–1068. https://doi.org/10.1021/acsphotonics.0c01600.

(17) Djorović, A.; Meyer, M.; Darby, B. L.; Le Ru, E. C. Accurate Modeling of the Polarizability of Dyes for Electromagnetic Calculations. *ACS Omega* **2017**, *2* (5), 1804–1811. https://doi.org/10.1021/acsomega.7b00171.

(18) Arp, T. B.; Kistner-Morris, J.; Aji, V.; Cogdell, R.; van Grondelle, R.; Gabor, N. M. Quieting a Noisy Antenna Reproduces Photosynthetic Light Harvesting Spectra. *Science (80-. ).* **2020**, *368*, 1490–149.

(19) Gentile, M. J.; Núñez-Sánchez, S.; Barnes, W. L. Optical Field-Enhancement and Subwavelength Field-Confinement Using Excitonic Nanostructures. *Nano Lett.* **2014**, *14* (5), 2339–2344. https://doi.org/10.1021/nl404712t.

(20) Núnez-Sánchez, S.; Lopez-Garcia, M.; Murshidy, M. M.; Abdel-Hady, A. G.; Serry, M.; Adawi, A. M.; Rarity, J. G.; Oulton, R.; Barnes, W. L. Excitonic Optical Tamm States: A Step toward a Full Molecular-Dielectric Photonic Integration. *ACS Photonics* **2016**, *3* (5), 743–748. https://doi.org/10.1021/acsphotonics.6b00060.

(21) Cleary, L.; Chen, H.; Chuang, C.; Silbey, R. J.; Cao, J. Optimal Fold Symmetry of LH2 Rings on a Photosynthetic Membrane. *Proc. Natl. Acad. Sci. U. S. A.* **2013**, *110* (21), 8537–8542. https://doi.org/10.1073/pnas.1218270110.

(22) Mustardy, L.; Garab, G. Granum Revisited . A Three-Dimensional Model – Where Things Fall into Place. *Trends Plant Sci.* **2003**, *8* (3), 117–122. https://doi.org/10.1016/S1360-1385(03)00015-3.

(23) Xiche, H.; Ritz, T.; Ana, D.; Felix, A.; Schulten, K. *Photosynthetic Apparatus of Purple Bacteria*; 2002; Vol. 35. https://doi.org/10.1201/9781351242899-6.

(24) Yang, Y.; Kravchenko, I. I.; Briggs, D. P.; Valentine, J. All-Dielectric Metasurface Analogue of Electromagnetically Induced Transparency. *Nat. Commun.* **2014**, *5*, 1–7. https://doi.org/10.1038/ncomms6753.

(25) Meinzer, N.; Barnes, W. L.; Hooper, I. R. Plasmonic Meta-Atoms and Metasurfaces. *Nat. Photonics* **2014**, *8* (12), 889–898. https://doi.org/10.1038/nphoton.2014.247.

(26) Hecht, M.; Wu, F. Supramolecularly Engineered J - Aggregates Based on Perylene Bisimide Dyes. *Acc. Chem. Res.* **2020**. https://doi.org/10.1021/acs.accounts.0c00590.

(27) Kriete, B.; Lüttig, J.; Kunsel, T.; Malý, P.; Jansen, T. L. C.; Knoester, J.; Brixner, T.; Pshenichnikov, M. S. Interplay between Structural Hierarchy and Exciton Diffusion in Artificial Light Harvesting. *Nat. Commun.* **2019**, *10* (1), 1–11. https://doi.org/10.1038/s41467-019-12345-9.

(28) Würthner, F.; Kaiser, T. E.; Saha-Möller, C. R. J-Aggregates: From Serendipitous Discovery to Supramolecular Engineering of Functional Dye Materials. *Angew. Chemie Int. Ed.* **2011**, *50* (15), 3376–3410. https://doi.org/10.1002/anie.201002307.

(29) Yakubovsky, D. I.; Arsenin, A. V.; Stebunov, Y. V.; Fedyanin, D. Y.; Volkov, V. S. Optical Constants and Structural Properties of Thin Gold Films. *Opt. Express* **2017**, *25* (21), 25574. https://doi.org/10.1364/oe.25.025574.

(30) Schnepf, M. J.; Mayer, M.; Kuttner, C.; Tebbe, M.; Wolf, D.; Dulle, M.; Altantzis, T.; Formanek, P.; Förster, S.; Bals, S.; König, T. A. F.; Fery, A. Nanorattles with Tailored Electric Field Enhancement. *Nanoscale* **2017**, *9* (27), 9376–9385. https://doi.org/10.1039/c7nr02952g.

(31) Brédas, J.-L.; Sargent, E. H.; Scholes, G. D. Photovoltaic Concepts Inspired by Coherence Effects in Photosynthetic Systems. *Nat. Mater.* **2017**, *16* (1), 35–44. https://doi.org/10.1038/nmat4767.

(32) Feist, J.; Garcia-Vidal, F. J. Extraordinary Exciton Conductance Induced by Strong Coupling. *Phys. Rev. Lett.* **2015**, *114* (19), 1–5. https://doi.org/10.1103/PhysRevLett.114.196402.

(33) Gonzalez-ballestero, C.; Feist, J.; Moreno, E.; Garcia-vidal, F. J. Harvesting Excitons through Plasmonic Strong Coupling. No. 1, 1–7.

(34) Şener, M.; Strümpfer, J.; Hsin, J.; Chandler, D.; Scheuring, S.; Hunter, C. N.; Schulten, K. Förster Energy Transfer Theory as Reflected in the Structures of Photosynthetic Light-Harvesting Systems. *ChemPhysChem* **2011**, *12* (3), 518–531. https://doi.org/10.1002/cphc.201000944.

(35) Koepke, J.; Hu, X.; Muenke, C.; Schulten, K.; Michel, H. The Crystal Structure of the Light-Harvesting Complex II (B800-850) from Rhodospirillum Molischianum. *Structure* **1996**, *4* (5), 581–597. https://doi.org/10.1016/S0969-2126(96)00063-9.

(36) Liu, L. N.; Sturgis, J. N.; Scheuring, S. Native Architecture of the Photosynthetic

Membrane from Rhodobacter Veldkampii. *J. Struct. Biol.* **2011**, *173* (1), 138–145. https://doi.org/10.1016/j.jsb.2010.08.010.

# Supplementary Information

## Bio-inspired polymers with polaritonic properties from VIS to IR: a material playground to mimic purple bacteria light-harvesting resonators.

Samuel Thomas Holder[1], Carla Estévez-Varela[2], Isabel Pastoriza-Santos[2], Martin Lopez-Garcia[3], Ruth Oulton[1] and Sara Núñez-Sánchez[2]

1 Quantum Engineering Technology Labs, University of Bristol, Bristol, UK

2 CINBIO, Universidade de Vigo, 36310 Vigo, Spain

3 International Iberian Nanotechnology Laboratory, Braga, Portugal

### 1.-Estimation of cyanine conformal structure on high concentrated J-aggregate:PVA solutions

The resulting dye and PVA mixtures before spin coating are too strongly absorbing to be able to measure the extinction in a cuvette without saturation, even with a path length of 1mm. In order to identify the supramolecular conformal structure of the self-assembled dyes, we measure the absorbance of a thin liquid layer of J-aggregate:PVA mixtures suspended between two glass slides pressed together thanks to two clips (figure 3.5). The absorbance is estimated using a spectrophotometer (Cary 4000 UV-Vis). In order to compare the spectral features with the absorbance of the dyes in solution, all the spectra were normalized to one to avoid effects associated to different optical paths due to sample preparation.

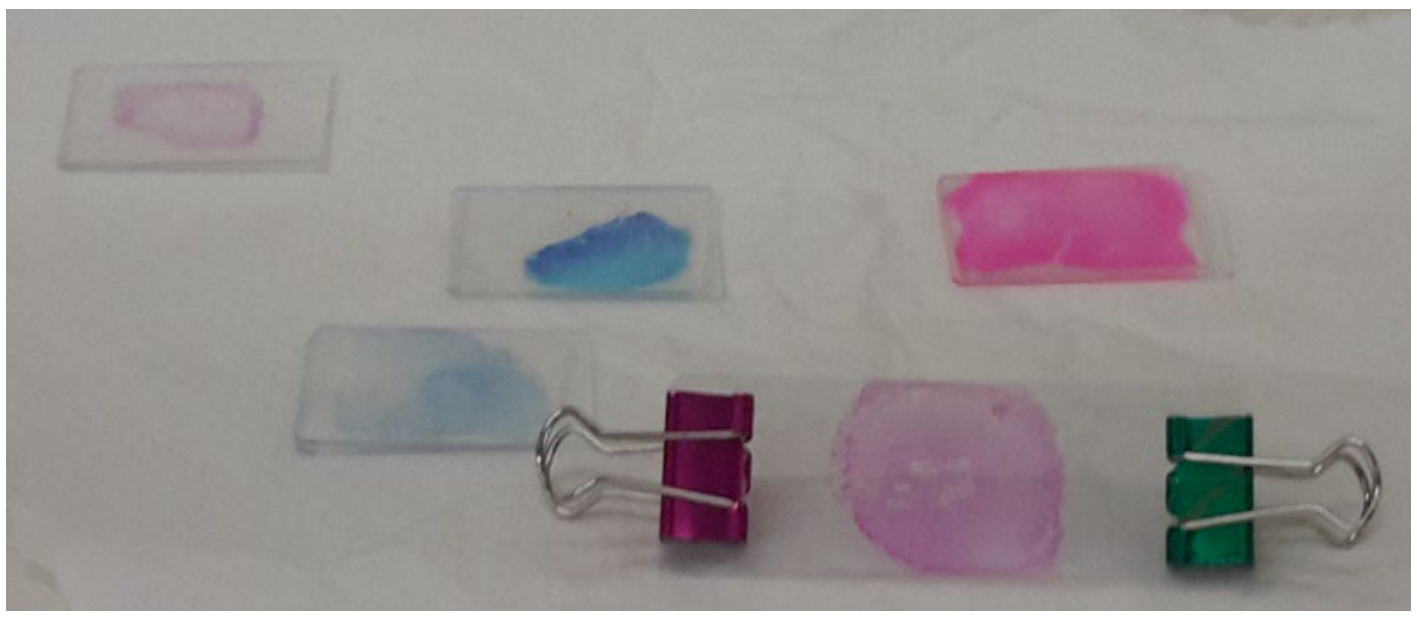

**Figure S1.** Picture of the J-aggregate:PVA solutions sandwiched between two microscope slides prepared to measure the extinction of the solutions in the spectrometer.

## 2.- AFM measurements and analysis

### 2.1.- Determination of thin film thickness.

The thickness of the samples was determined by Atomic Force Microscopy (AFM, Agilent 5420 in tapping mode). Samples were scratched in a line using a flat clean room tweezer in several positions along with the sample. The thickness is estimated from the difference between the channel and the top of the film. Choosing two strips rather than averaging the whole image has two advantages: anomalous features such as dirt can be avoided, and the levelling of the image can be checked: both regions should give the same film thickness. Within these strips, which run left to right and are marked by white dotted lines in Figure S2, areas corresponding to the bare substrate and within 10μm the step edge are chosen by eye, and the average height of these areas is calculated. This gives an estimate of the substrate height. The same method is applied to the film area within the same strip to give an estimate of the film height. The difference in these average height values gives an estimate of film thickness, which is reported in table 1. The error on this thickness estimate is calculated by considering the independent height estimates that are averaged to estimate the substrate and film heights. The uncertainty in the film thickness estimate is given by the variance of these mean values. The standard deviation of the mean of N independent samples, $\sigma_N$, for samples from a normal distribution with standard deviation $\sigma$, is

$$\sigma_N = \frac{\sigma}{\sqrt{N}}.$$

Not every measured point in these AFM images can be considered an independent height sample. The scanning of the atomic force microscope tip across the sample introduces correlations between subsequent height samples. Here these correlations have a characteristic length of around 1μm or five pixels. To get a lower bound on the number of independent height samples measured, N, we divided the total number of averaged points by 25, to impose a 1μm square pixel size. These N values together with the measured standard deviation of heights ($\sigma$), for substrate and film areas, gives the uncertainty in the height estimate assigned to each area via equation 3.2. These uncertainties are combined to give the uncertainty of the thickness estimate (table 1).

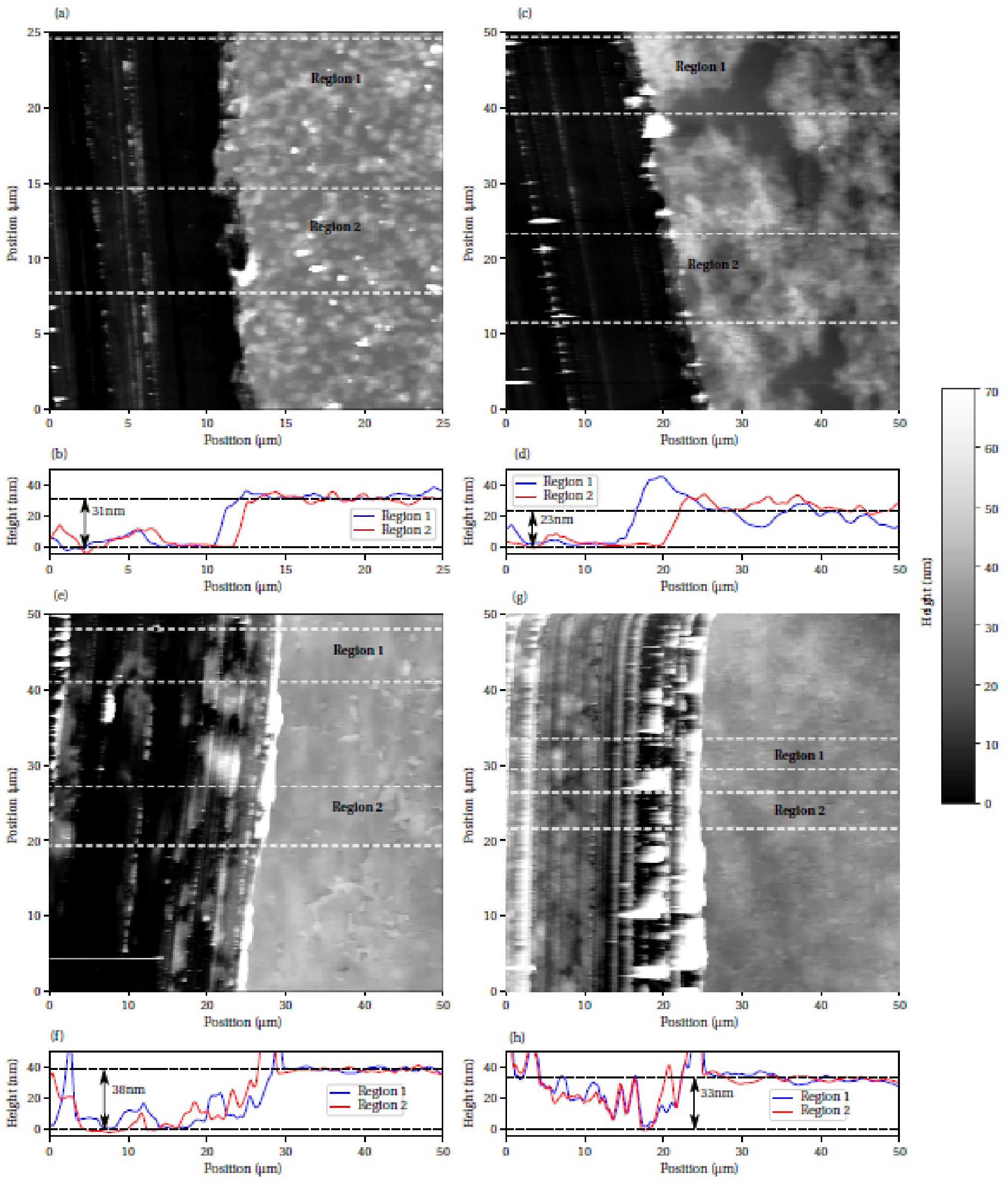

**Figure S2:** Atomic force microscope images of a substrate to film step in our J-aggregate and PVA films: (a) J562, (b) J587, (c) J619 and (d) J798. Each AFM image is marked with regions 1 and 2: the average profile of each of these strips is plotted below each AFM image. The substrate and film heights are marked with dotted red lines. Note that the J562 image is of a smaller (25µmby 25µm) area.

### 2.2.- Determination of surface roughness.

Surface roughness can modify the optical response of a material. Surface roughness can be quantified as the standard deviation (σ) of the distribution of measured surface heights (S):

$$\sigma^2 = \langle(\langle S\rangle - S)^2\rangle$$

Note that σ is the conventional notation for the standard deviation of the measured surface heights, or equivalently the root-mean-square roughness. The standard deviation σ quantifies the width of the distribution of surface heights. Note that this simple metric does not use information about the height difference between adjacent points on the surface, and hence surfaces with very different textures can have the same roughness. However, $\sigma$ is useful for estimating how big an effect the surface roughness will have on the optical response of a thin film. In table 1 we report the roughness of each of our J-aggregate and PVA thin films, based on the figure S3 images where the pixel spacing (10nm) and the tip radius (<10nm) are comparable, meaning that the measured height image is close to the actual surface profile with only features below 10nm size being smoothed out. The statistical uncertainty of the measured $\sigma$ values is given by

$$\sigma_\sigma^2 = \sigma \frac{1}{\sqrt{N}}$$

where N is the number of independent height samples used to calculate $\sigma$. Here, N is not identical to the number of pixels in the image. Instead, independent height samples are those separated by at least the correlation length of the surface.[1]

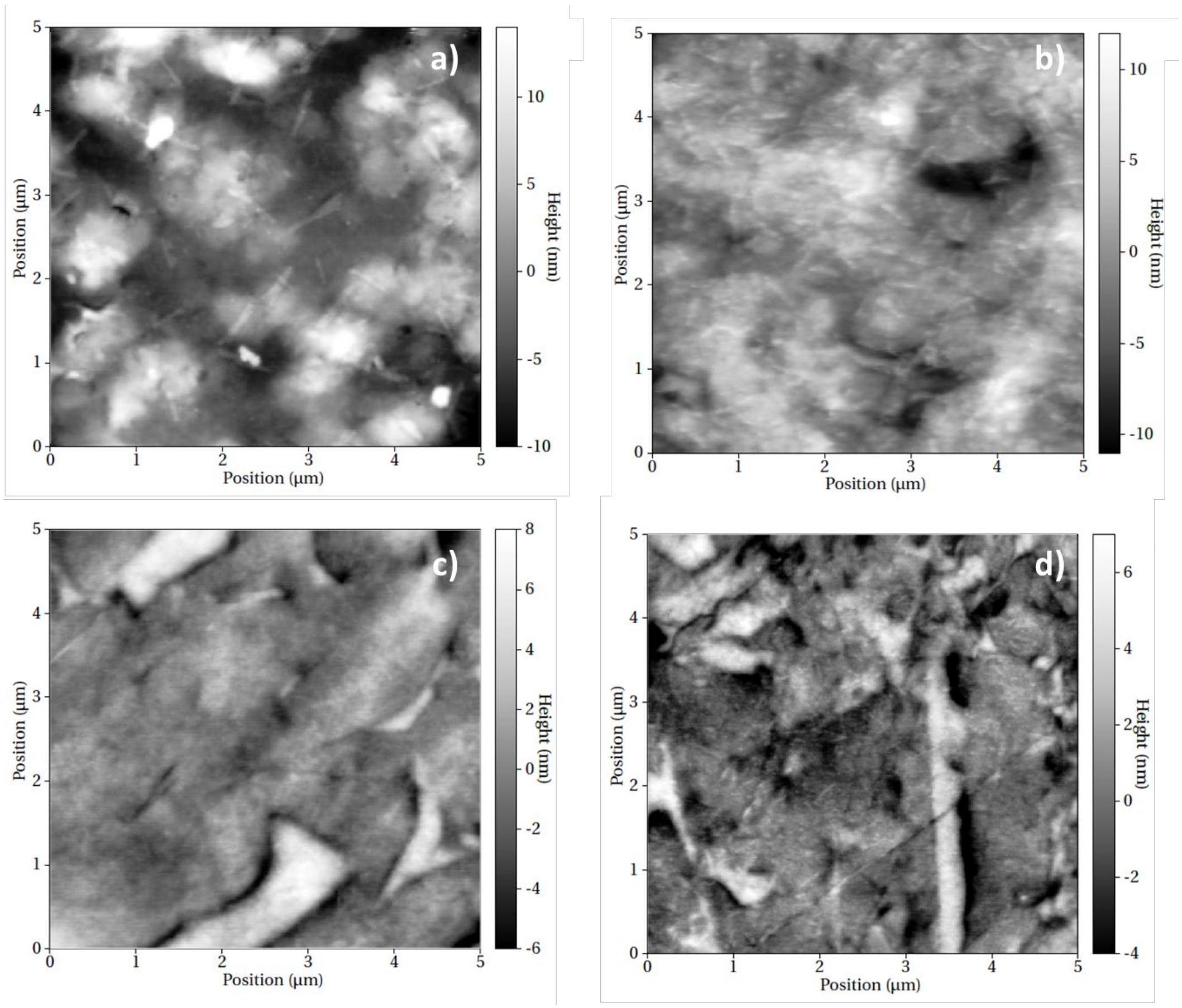

**Figure S3**. Height AFM images of the J-aggregate:PVA films of (a) J562, (b) J587, (c) J619, (d) J798 dye.

### 3.- Fourier imaging spectroscopy under Kretschmann prism-coupling configuration.

A microscope cannot be used with a prism to make an angular reflectance measurement, as the short focal length objective lens requires for the sample to be brought close to the lens focussed on its surface. However, the Kretschmann prism-coupling configuration can be realised without a prism using an oil-immersion objective lens and a thin film sample deposited on a glass microscope coverslip (figure S4). In this implementation, the high index region from which light is incident on the sample is formed by three different media: the bottom spherical lens of the oil-immersion objective, the immersion oil, and the glass substrate.

The samples were under critical illumination by a fibre-coupled tungsten lamp giving broadband illumination. The Fourier plane of the oil-immersion objective was imaged onto the tip of a scanning fibre which scans the Fourier plane along a straight line intersecting the centre of the Fourier plane, with each position corresponding to a different angle of incidence. The fibre output was coupled to a spectrometer allowing a broad wavelength range to be measured at each angle of incidence. The measured polarization was determined by the relative orientation of the measured line on the Fourier plane and a polarizer in the illumination path, to obtain either the p- or s-polarised reflectance.

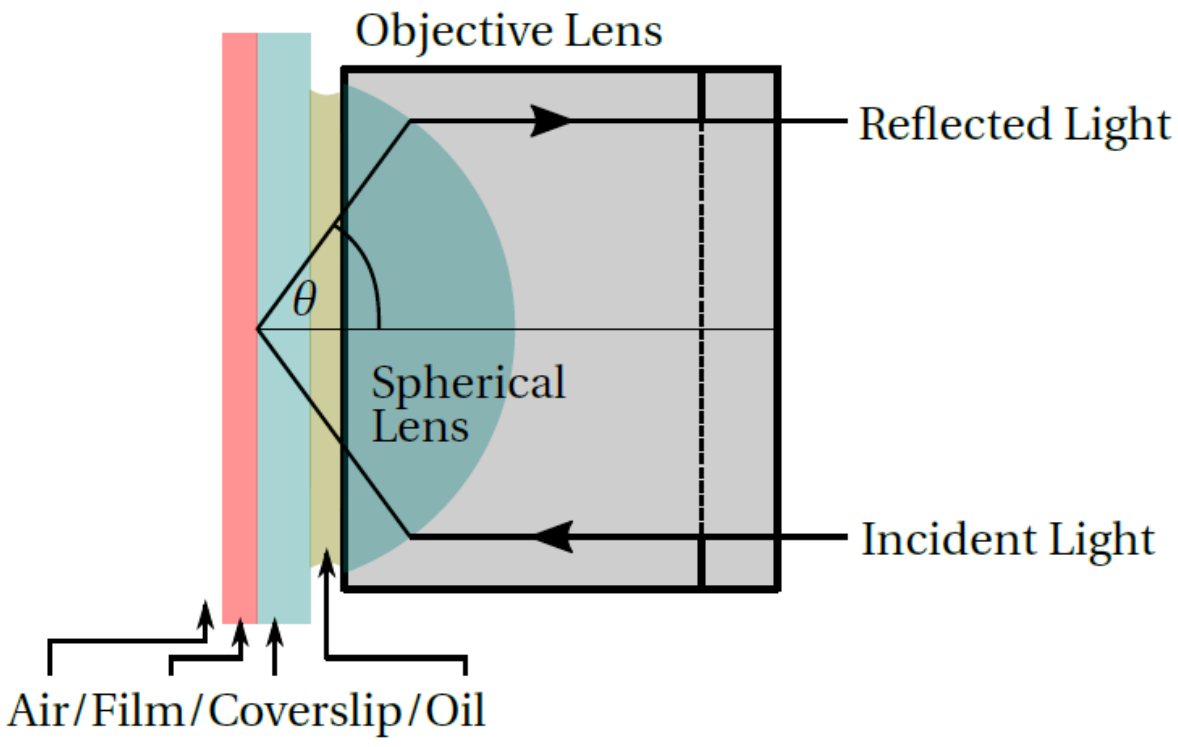

**Figure S4:** Scheme of the implementation of the Kretschmann prism-coupling configuration using an oil immersion objective lens.

To calibrate the angles on the image of the k-space obtained in the back focal plane of the objective, we used a coverslip as equivalent to the response of a prism with the refractive index of the glass (1.46). For angles of incidence smaller than the critical angle of a glass prism almost all the light pass through the interface glass/air and we should observe a minimum in reflectance. However, for angles larger that the critical angle, all the light is reflected due to matching total internal reflection conditions. Therefore, in the k-space, the response of a prism should be a black circle whose radius is defined by the total internal reflection condition (see Figure S5). The critical angle occurs when the x-component of the wavevector of light in the prism can no longer be matched to the x-component of the wavevector of light in the air above the prism. Moreover, to take a spectra reference of the source, we used a silver mirror as a reference with a flat reflection band in the visible with a reflectance close to a 100%. Figure S5 shows the reflectance obtained from a cover glass (equivalent of a prism) calibrated for the (a) scanning fibre.

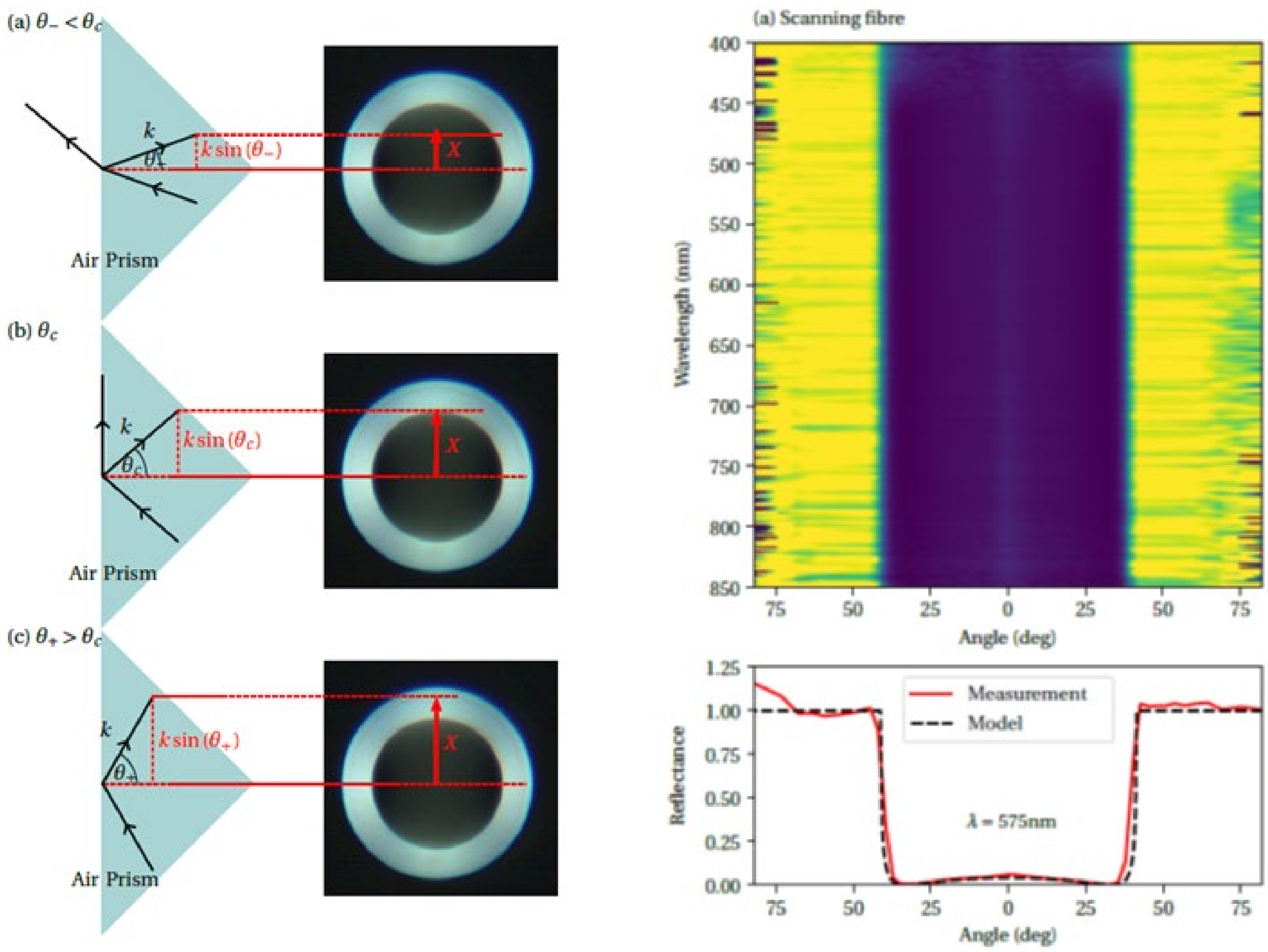

**Figure S5:** Reflectance from a prism at a range of angles, and the corresponding points on the observed camera image of the Fourier plane at three different angles: (a) below the critical angle, (b) at the critical angle and (c) above the critical angle. Reflectance of p-polarised light from a glass coverslip in the Kretschmann configuration, using different spectroscopic detection methods: (a) scanning fibre. The bottom panels show a cross-section through the above images at wavelength 575nm.

**5.-Defined objects and boundary conditions on the electromagnetics simulation set-up.**

The optical response of nanoring structures with external diameter of 9 nm, an inner diameter of 3.1 nm and a height of 5 nm were obtained by FDTD simulations using commercial Lumerical software (Figure S6.a). These simulations have been performed for five different nanorings. Four of them will be composed by the optical properties of the four J-aggregate:PVA polymers. The fifth nanoring structure will be composed by just a dielectric polymeric matrix with a constant refractive index of 1.46 like PVA. This will be the reference for a LH2 ring architecture composed just by proteins without any chromophore. In all the cases the surrounded medium is water with a refractive index of 1.33.

**5.1.-Estimation of absorption, scattering and extinction cross-sections for non-symmetric three-dimensional (3D) objects.**

The absorption and scattering cross-sections were estimated by the combination total-field scattered-field (TFSF) source which injects a linear polarized plane wave with a finite span and a cross-section analysis group. The cross-section analysis group estimated the net power flowing through a 3D-field monitor with a box shape enclosing the nanoring normalized to the source intensity. In the case of the cross-section analysis group located inside the TFSF source we obtained the absorption cross-section and, in the case of the cross-section analysis group located outside the TFSF source we obtained the scattering cross-sections. All data is normalized to nanoring surface to obtain the cross-section efficiency per particle (equation S1).

$$\sigma_{particle} = \sigma_{total}/A_{ring}, \text{ equation S1.}$$

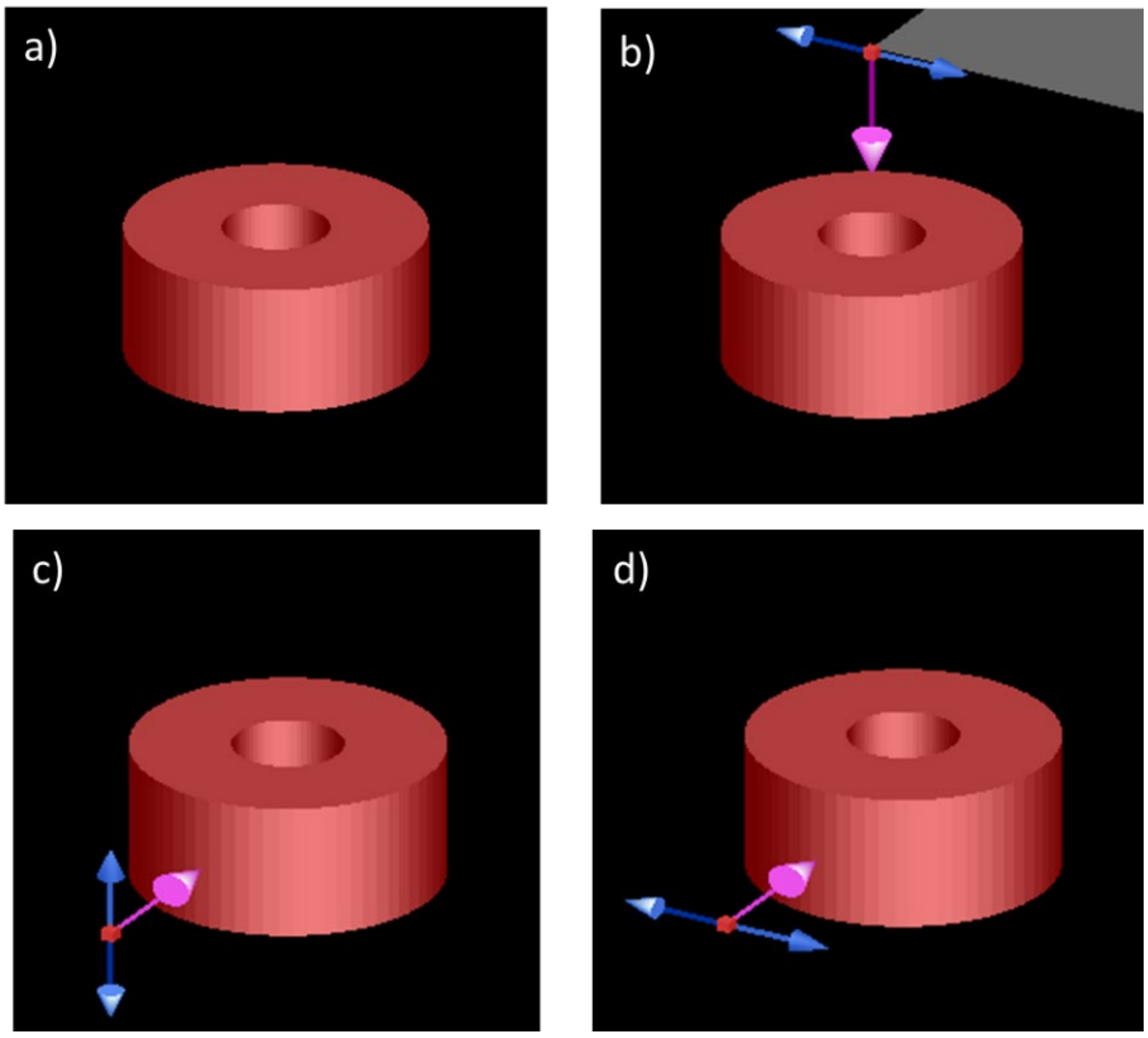

**Figure S6.** a) Simulation object created to mimic the LH2 ring architecture of lamellar membranes of Rhodospirillum molischianum with an external diameter of 9 nm, an inner diameter of 3.1 nm and a height of 5 nm. Incident light was set as linear polarized total-field scattered-field (TFSF) sources. The pink arrow indicates the propagation direction (wavevector) while the blue arrows indicate the direction of the polarization vibration. Due to the geometry of the ring architectures three polarizations (a – polarization 1, b – polarization 2 and c-polarization 3) were set up to cover all potential combinations object-polarization in order to determine the total absorption, scattering and extinction cross sections for un-polarized light (as under sun illumination).

Because the cross sections are associated to three dimensional (3D) calculations, the relative orientation of incident polarized light with the object can play a role in how light is absorbed and scattered. In the case of spherical nanoparticles, the object is symmetric in the three dimensions, therefore it can be estimated by just by the cross sections for one polarization. However, the nanorings have a rotational symmetry in the vertical direction as cylinders. Therefore, the total cross sections were calculated considering all possible relative orientations of incident light and the nanorings. Figure S6.b-c shows the three different polarizations that we have considered for our calculations. The total cross sections were established by the average of all polarizations considering all the potential orientations and symmetries within the 3D-field monitor box (Figure S6, equation S2).

$$\sigma_{total} = \sigma_{polarization1} * 2\,(sides) + \sigma_{polarization2} * \frac{4\,(sides)}{2\,(polarizations)} + \sigma_{polarization3} * \frac{4\,(sides)}{2\,(polarizations)}$$ , equation S2.

Figure S7 shows the total absorption and the total scattering cross sections for the four J-aggregate:PVA nanorings and the reference nanoring without chromophores. The reference nanoring shows no-absorption and a flat extinction cross. However, the J-aggregate:PVA nanorings shown a clear absorption and scattering peaks at shorter wavelengths than the absorption of the bulk material which can be associated to a local surface exciton resonances.

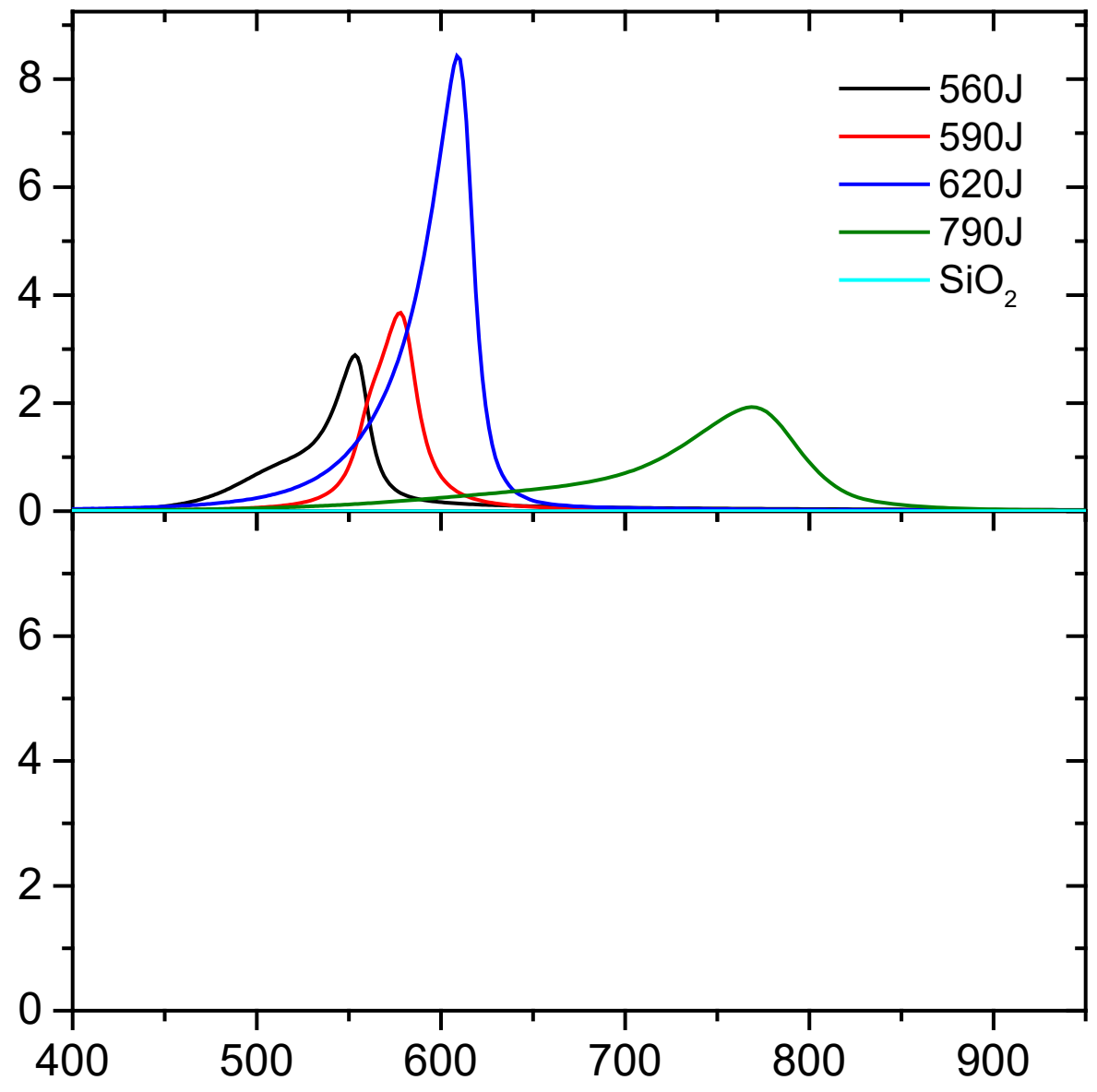

**Figure S7.** a) Absorption and b) scattering cross sections obtained for nanorings mimicking LH2 ring architecture of lamellar membranes of *Rhodospirillum molischianum* with the optical properties obtained for the bulk J-aggregate polymer films. Absorption cross-sections are peaked at 554, 578, 609 and 770 nm for J562, J587, J619 and J798 dyes, respectively. Scattering cross sections are peaked at 556, 580, 612 and 780 nm for J562, J587, J619 and J798 dyes, respectively.

The results shown in Figure 4.b shows the total extinction efficiency which is the extinction cross section normalized by the nanoring area (see equation S3).

$\sigma_{extinction,\ total} = \sigma_{absorption,total} + \sigma_{scattering,total} \Rightarrow Q_{extinction} = \sigma_{extinction,total} / Area_{nanoring}$ , equation S3.

**5.2.- Local electric field distribution for perpendicular illumination to the lamellar membrane and polarization contained in the plane.**

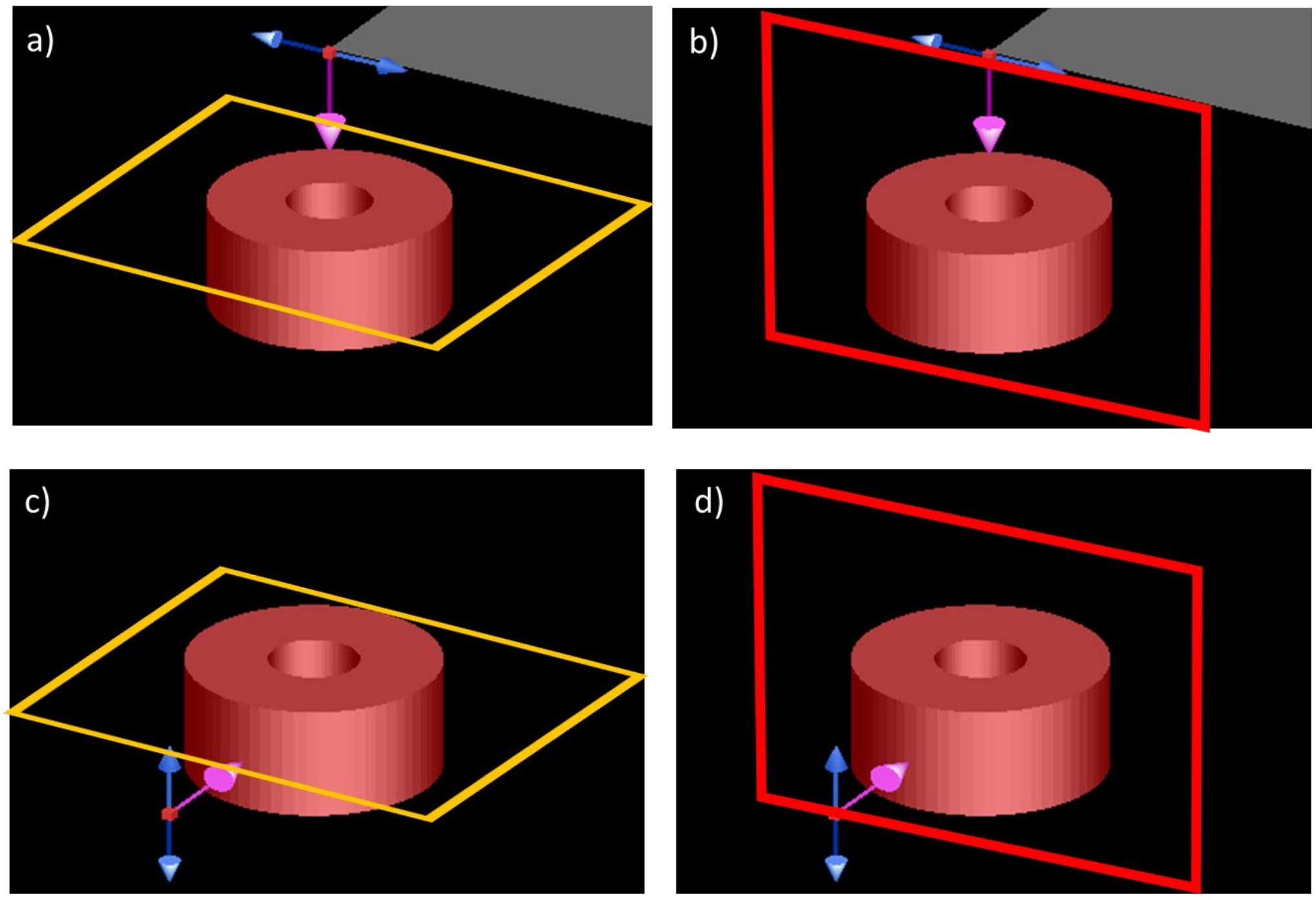

**Figure S8.-** a-b) Configuration of simulation with the light travelling perpendicular to the membrane. c-d) Configuration with the light travelling parallel to the membrane. The red and yellow squares are the planes where we have estimated the distribution of the local field intensities.

LH1 and LH2 ring complexes are ring architectural structures supported by a lamellar membrane located at the bottom of XY plane of the Figure S8. The yellow and red rectangle are the two planes where we have simulated the electric field intensity distribution for all the nanorings.

**5.3.- Local electric field distributions for J562, J587 and J798 dyes at LSER peaks.**

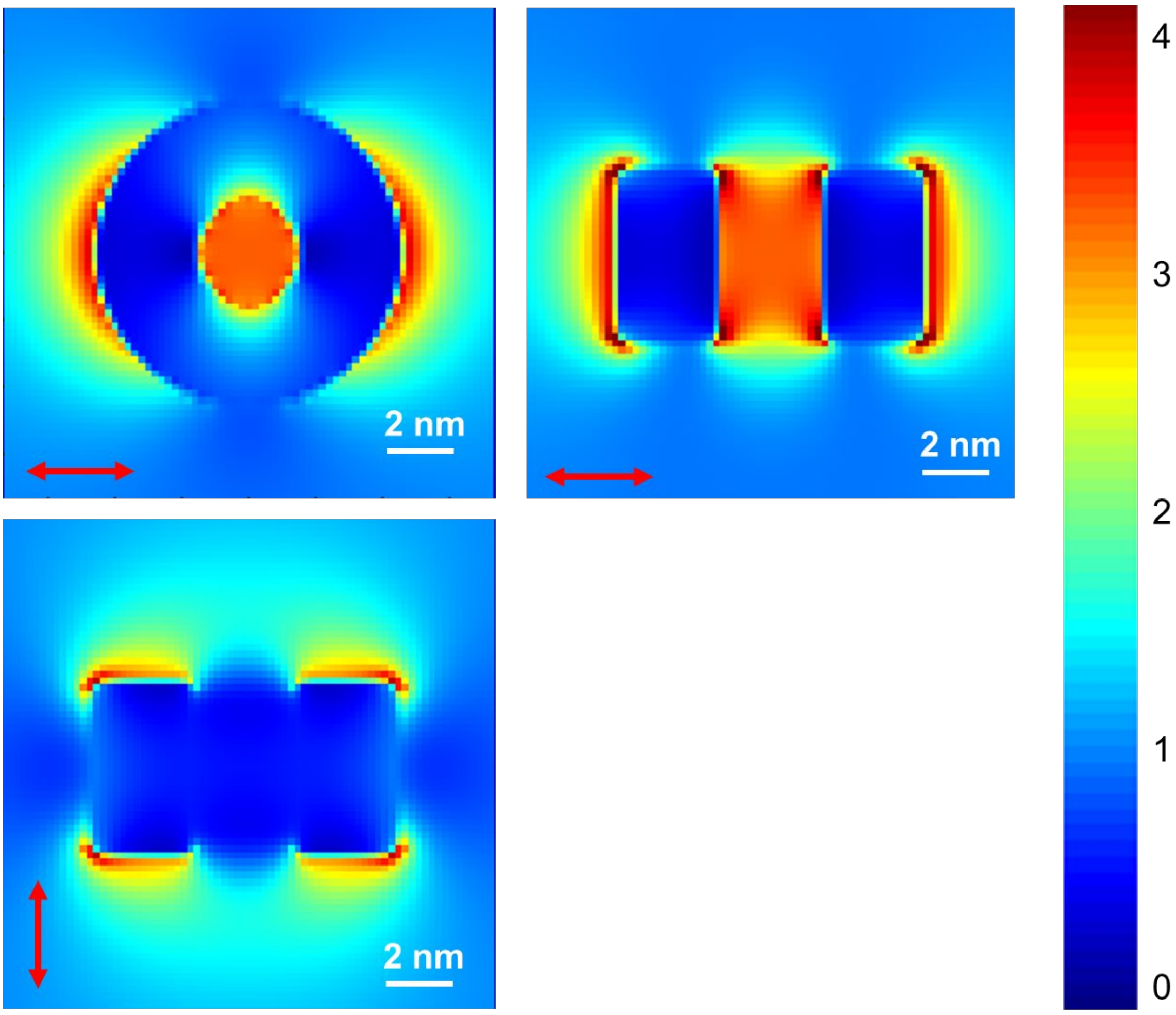

**Figure S10.-** Electric field intensity distribution for a nanoring composed by J-aggregate:PVA material of J562 dye at the LSER peak at 553 nm. Light linear polarized travelling towards the plane. Polarization indicated by red arrows.

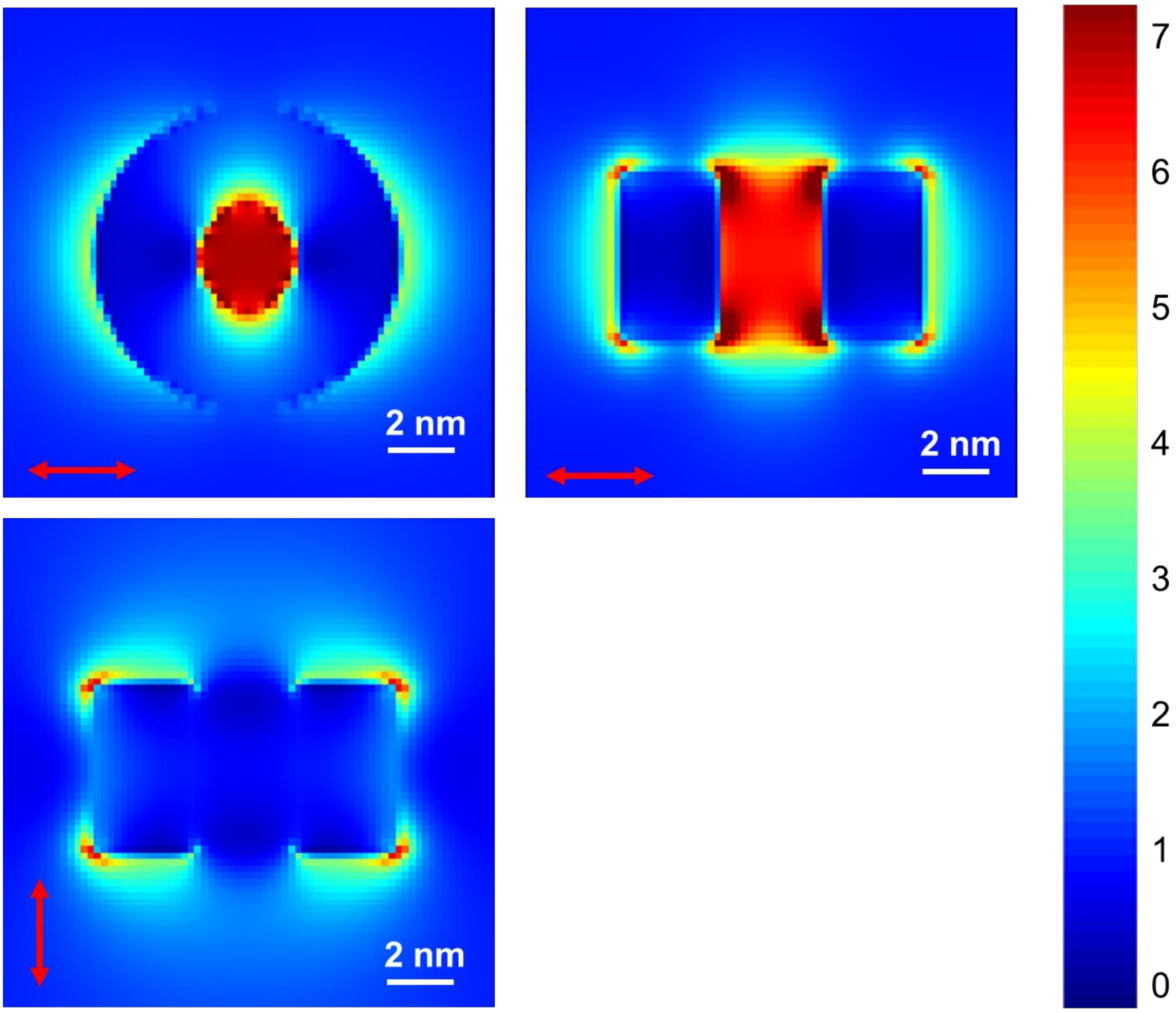

**Figure S11.-** Electric field intensity distribution for a nanoring composed by J-aggregate:PVA material of J587 dye at the LSER peak at 578 nm. Light linear polarized travelling towards the plane. Polarization indicated by red arrows.

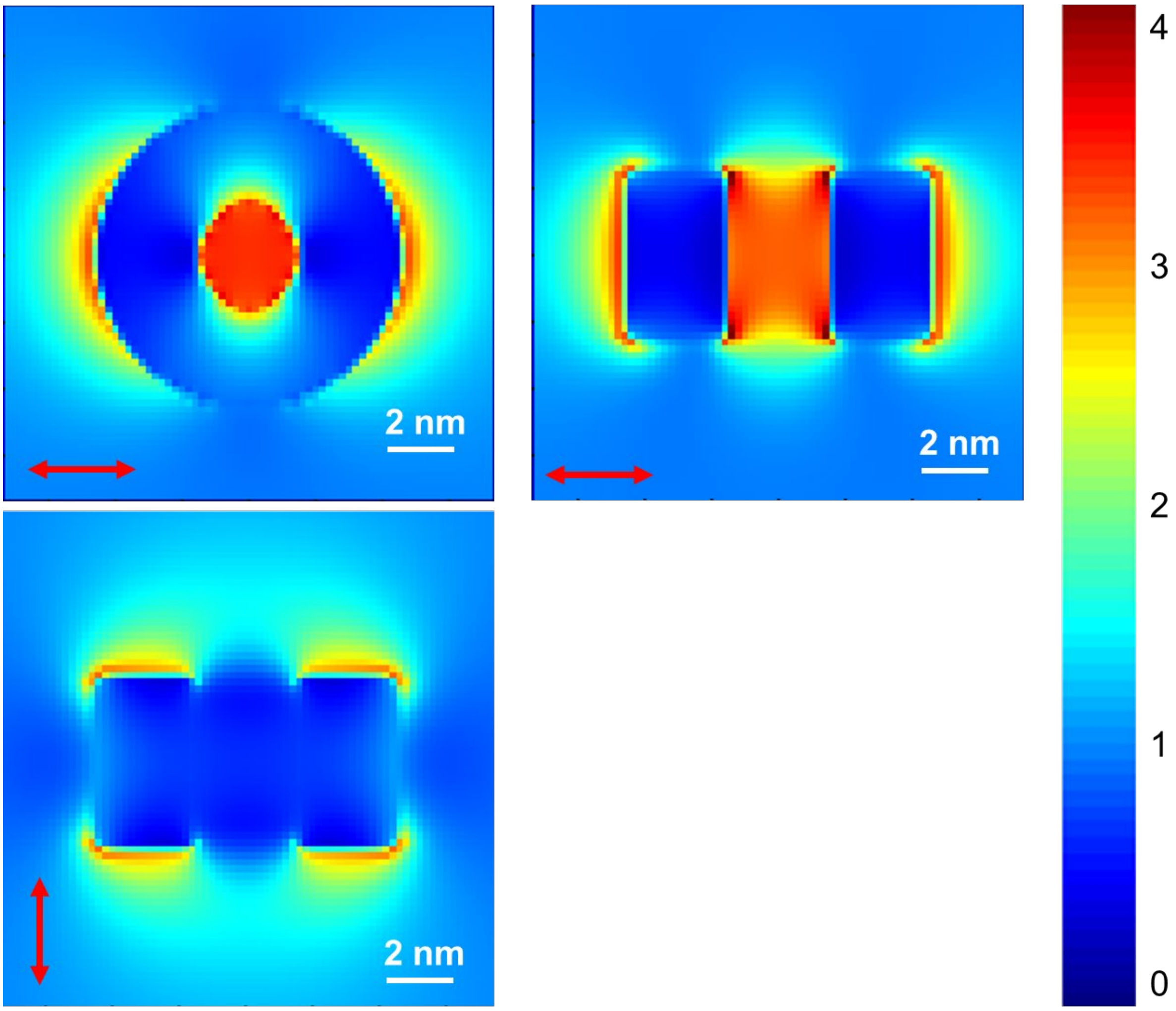

**Figure S12.-** Electric field intensity distribution for a nanoring composed by J-aggregate:PVA material of J798 dye at the LSER peak at 768 nm. Light linear polarized travelling towards the plane. Polarization indicated by red arrows.

**REFERENCES**

(1) Mack, C. A. Uncertainty in Roughness Measurements: Putting Error Bars on Line-Edge Roughness. *J. Micro/Nanolithography, MEMS, MOEMS* **2017**, *16* (1), 10501. https://doi.org/10.1117/1.JMM.16.1.010501.